\documentclass[aps,twocolumn,english]{revtex4}

\usepackage{epsfig}
\usepackage{graphicx}
\usepackage{amsmath}
\usepackage{babel}

\begin{document}
\makeatother


\title{Neutron capture cross sections of radioactive nuclei}
\author{C. A. Bertulani}
\affiliation{Department of Physics and Astronomy, Texas A\&M University-Commerce, Commerce, TX, USA}
\email{carlos.bertulani@tamuc.edu}
\author{B. V. Carlson}
\affiliation{Departamento de F\'{\i}sica, Instituto Tecnol\'ogico de Aeron\'autica-CTA, 12228-900, S\~ao Jos\'e dos Campos, Brazil}

\date{\today}


\begin{abstract}
{\bf Abstract:}
Alternative methods to calculate neutron capture cross sections on radioactive nuclei  are reported using the theory of Inclusive Non-Elastic Breakup (INEB) developed by Hussein and McVoy \cite{HM1985}. The statistical coupled-channels theory proposed in Ref. \cite{BDH14} is further extended in the realm of random matrices.  The case of reactions with the projectile and the target being two-cluster nuclei is also analyzed and applications are made for  scattering from a deuteron target \cite{Hus20}. An extension of the theory to a three-cluster projectile incident on a two-cluster target is also discussed. The theoretical developments  described here should open new possibilities to obtain information on the neutron capture cross sections of radioactive nuclei using indirect methods.
\end{abstract}

\maketitle

{{\bf keywords:} neutron capture, radioactive nuclei, inclusive reactions}

\bigskip 

{\it To the Memory of Mahir Hussein}

\section{Introduction} 
Research on reactions involving radioactive nuclei has provided invaluable information about the properties of nuclei close to the drip line \cite{BCH93,BG10,AB20}. Further, indirect methods with rare isotopes yield cross sections of neutron capture  and of other fusion reactions needed to fill the knowledge gaps in reaction chains such as the r- and s-processes of interest for astrophysics \cite{CK16}. The experimental information on neutron capture reactions is mostly constrained by capture on stable nuclei. The neutron capture on radioactive nuclei, in particular those near the drip nuclei, are difficult to obtain and are not available or have to be inferred indirectly \cite{BG10,Trib14}. Another way to obtain neutron capture cross sections on radioactive nuclei is by means of indirect hybrid reactions. Among others, a few methods used frequently by experimentalists are the surrogate method \cite{ Escher2012,Esch18,potel2017}, the Trojan horse method \cite{Bau86,Spit01,Tum16,Hus17,BHT18}, or the ANC method \cite{MT90,Xu94,Muk01}. The surrogate method is most often used to infer neutron capture cross sections for fast neutrons on actinide nuclei with the purpose to study fast breeder reactors. In the Trojan horse method one is interested to extract information on the reaction $x+A\rightarrow y+B$ by studying a reaction of the form $a+A\rightarrow b+y+B$. The ANC method uses transfer reactions to extract asymptotic normalization coefficients useful to calculate radiative capture reactions of the form $A + a \rightarrow B + \gamma$. All these indirect methods have one thing in common:  the many-cluster feature of nuclear reactions. 

The method employed in this article is the Inclusive Nonelastic Breakup  (INEB) reaction theory initially developed in Refs. \cite{IAV1985,UT1981,Ichimura1982,HM1985} and widely popularized by Hussein and collaborators (see, e.g., Ref. \cite{Hu17}). In this article we also discuss the use of the INEB theory to study the case of breakup reactions of a radioactive projectile on a deuteron target, including a further extension of the method employed in Ref. \cite{Bert19}. The INEB theory can be easily applied to the case of a long lived radioactive target, such as $^{135}$Xe. The theory is also adaptable to study transfer in reactions involving multi-clusters both in the projectile and the target, starting with the simpler (d,p) case (a modern example is Ref. \cite{Pai15}). This is often the case in transfer reactions with rare nuclear isotopes. For the simpler cases of a weakly-bound projectile or for the coupling to compound nucleus states  too many channels might be involved, with computations costing prohibitive large CPU times. An statistical theory to tackle these cases have been proposed in Ref. \cite{BDH14}. In this article we show that another path to simplify the theoretical description of the reaction mechanism can be achieved by using statistical concepts  involving random matrices. The INEB theory is relatively simple since reaction mechanisms can be quite difficult to describe theoretically. It is interesting to note  that one of the first publications concerning reactions involving unstable targets was done by Chew and Low back in 1959 \cite{CL1959}. Nowadays, with the construction of radioactive beam facilities, the most intense experimental studies involve radioactive projectiles reacting with stable nuclear targets.

For the extension of the INEB theory to multi-cluster reaction cases, we consider first the case of a three-body non-cluster radioactive projectile reacting with a two-body cluster target, using the deuteron as a surrogate. The reaction in this case is a neutron pickup reaction. The main idea of the surrogate method is that a measurement of the inclusive proton recoil spectrum allows the extraction of quantities involved in neutron capture reactions \cite{ Escher2012,Esch18,potel2017}. But, the capture reaction is not the free neutron capture. Different reaction mechanisms are involved in the surrogate reaction owing to the fact that the neutron within the deuteron is bound and carries the details of its internal wavefunction.  We also consider a second case in which a four-body reaction involves a two-fragment projectile and a two-fragment target. As examples of  these reactions we mention the one proton halo nucleus $^8$B,  involved in the reactions $^8$B + d $\rightarrow$ p + $^9$B or p + ($^7$Be + d). Thus, two features will be exhibited in the inclusive proton spectrum: (a) at low proton energies the incomplete fusion of $^7$Be + d will play a role and (b) the capture neutron reaction will be influenced by higher proton energies. The case of one-neutron halo nuclei are then considered. $^{11}$Be and $^{19}$C are good projectile examples. As an example of an outcome of such studies, we anticipate that the inclusive proton spectrum would show a low energy peak at 2.22 MeV, and a weaker peak at a higher energy of $\approx$ 5.00 MeV \cite{Hus20}.

The developments reported here will be useful to test the applicability of the INEB theory to large neutron capture cross sections such as n+$^{135}$Xe where the $^{135}$Xe has a lifetime of 9.8 hours.   $^{135}$Xe is a notorious nuclear reactor poison and it is widely known that its thermal neutron capture cross section is incredibly large, of the order of $2.5\times 10^6$ barns. One of the goals of this theoretical exercise is to use  $^{135}$Xe  as a benchmark to assess how the proton spectrum emerges in, e.g., a reaction of the type d + $^{135}$Xe $\rightarrow$ p + $^{136}$Xe. Further, our multi-cluster discussion should constitute an important extension of the four-body formulation involving a three fragment projectile on a ``structureless" target developed recently in Refs. \cite{CFH2017,Bert19} to two-fragment on two-fragment nuclear reactions  \cite{Hus20}. 
It would be more convenient to use $^{10}$B as a target as it is stable. The thermal neutron capture in this case is about 4000.00 barns, very large considering the size of this nucleus and the fact that other nuclei in its vicinity have much smaller cross sections. This nucleus, is used in radiotherapy under the name Boron Neutron Capture Therapy. It would be interesting to extract the modified capture cross section $\hat{\sigma}$(n + $^{10}$B) and check wether it is feasible to use the inclusive proton spectrum in the breakup reaction d + $^{10}$B $\rightarrow$ p + (n + $^{10}$B) to obtain the capture cross section. This would supply a check on the consistency of the theory.
The surrogate method has also been benchmarked using the INEB theory in Refs. \cite{Potel2015, Moro2-2015,Moro2017, Ducasse2015,Carlson2015}. 

\section{Inclusive Non-Elastic Breakup Reactions of radioactive projectiles with two cluster targets}
\subsection{Radioactive projectiles and neutron poisons }
Let us take a radioactive nuclear poison projectile, e.g., $^{135}$Xe incident on a two-cluster target such as the deuteron. The lifetime of $^{235}$Xe is very long, about 9.6 hours, easily allowing the production of a radioactive beam with present techniques. We would like to describe a reaction of the type $^{135}$Xe + d $\rightarrow$ p + $^{136}$Xe, which is known as a pickup reaction. We will also assume that the proton energy spectrum is possible to be measured. We give a short summary of the theoretical equations developed in Refs. \cite{IAV1985,UT1981,Ichimura1982,HM1985} which are pertinent to the present discussion.  These theories would be appropriate to extract the neutron capture cross section n + $^{135}$Xe $\rightarrow$ $^{136}$Xe from the experimental analysis of the  $^{135}$Xe + d $\rightarrow$ p + $^{136}$Xe reaction.  This is a general reaction of the type a + A $\rightarrow$ p + (n + A) and the INEB formulation of Ref. \cite{HM1985}  yields for the differential cross section
\begin{equation}
\frac{d^{2}\sigma^{\rm INEB}_p}{dE_{p}d\Omega_{p}} = \hat{\sigma}_{R}^{n}  \ \rho_{p}(E_p, \Omega_p), \label{inebI}
\end{equation}
where  $\rho_{b}(E_p, \Omega_p)$ is the density of states of the spectator fragment, $p$, as measured in the experiment and where $\hat{\sigma}_{R}^{n}$ is the  total reaction cross section of the process n + A including medium corrections,
$
\hat{\sigma}_{R}^{n} = \hat{\sigma}_{R}(n + A).
$

 Therefore, in this theory, the breakup cross section is directly proportional to the reaction cross section $\hat{\sigma}_{R}^{n}$ obtained from
\begin{equation}
\hat{\sigma}_{R}^{n} = - \frac{k_n}{E_n}\langle\hat{\rho}_{n}(\textbf{r}_n)\left|W_{n}(\textbf{r}_n)\right|\hat{\rho}_{n}(\textbf{r}_n)\rangle  ,
\label{sigR}
\end{equation}
where a complex optical potential $U_{n}=V_n+iW_{n}$ is assumed for the interaction between the  neutron and the target A. Note that we avoid the practical discussion of how to theoretically obtain a consistent optical potential from first principles. Recent work based on the concept of self-energy \cite{BVG80} has recently been the focus of reaching a self-content theory for optical potentials including modern approaches of nuclear forces, such as chiral effective field theories \cite{Holt13}.  For a recent work in this direction, see Ref. \cite{Tayl19}.

The most difficult part of the reaction formalism is to calculate the ``source" function $\hat{\rho}_{n}(\textbf{r}_n)$ which is the overlap of the neutron distorted wave and the total wave function of the surrogate nucleus $d$ in the  incident channel. Eq. \eqref{sigR} is exact, but readily amenable to approximations. Using the DWBA limit and the post-form of the interaction, $V_{pn}$, one can show that \cite{HM1985} 
\begin{equation}
\hat{\rho}_n(\textbf{r}_n) = (\chi^{(-)}_{p}|\chi^{(+)}_{d}\Phi_{d}>(\textbf{r}_n) ,\label{rho1}
\end{equation}
where $\Phi_{d}(\textbf{r}_p, \textbf{r}_n)$ is the internal wavefunction of the $d=p+n$ system.
In Ref. \cite{IAV1985} a different approach was taken, based on the post form of the interaction $V_{np}$. In their case, the source function includes a Green's function accounting for the neutron propagator, namely, 
\begin{equation}
\hat{\rho}_n(\textbf{r}_n) = \frac{1}{E_n - U_n + i \varepsilon} (\chi^{(-)}_p|V_{np}|\chi^{(+)}_{d}\Phi_{d}\rangle \label{rho2} .
\end{equation}

As shown in Ref. \cite{Bert19}, if one additionally assumes that the three-body distorted waves $\chi_{i}$ depend only on the relative coordinates between the fragment and the target, the cross section in Eq. (\ref{sigR}) can be further decomposed into
\begin{equation}
\frac{d^{2}\sigma_{p}}{dE_{p}d\Omega_b} = \frac{E_{n}}{k_n} \int d\textbf{r}_{n} |\hat{S}_{p}(\textbf{r}_n)|^{2} W(\textbf{r}_n)|\chi^{(+)}_{n}(\textbf{r}_{n})|^2 ,
\label{sigRpar}
\end{equation}
where  
\begin{equation}
\hat{S}_{p}(\textbf{r}_n) \equiv \int d\textbf{r}_{p}  \langle\chi^{(-)}_{p}|\chi^{(+)}_{p}\rangle(\textbf{r}_p)\Phi_{d}(\textbf{r}_p, \textbf{r}_n).
\label{S}
\end{equation}

This formalism was the starting point of an analysis performed in Ref. \cite{Bert19} To study reactions involving the deuteron as a projectile and neutron poisons, such as $^{135}$Xe as a target or vice-versa. In any case, this surrogate reaction will certainly yield useful information on the total reaction cross section for n + $^{135}$Xe, as displayed in the equations above. To obtain the capture cross section one needs to take the difference between the reaction cross section and the other direct reaction contributions, as for example the inelastic excitation of $^{135}$Xe. 

Several neutron poisons exist, even stable nuclei such as $^{10}$B and $^{157}$Gd,  for which the cross sections for capture of epithermal  neutrons with $E_n \lesssim 0.4$ eV  are $3.80 \times 10^3$ barns and $2.54 \times 10^5$ barns, respectively. A very large ($8.61 \times 10^5$ barns) thermal neutron absorption cross section  was also observed for $^{88}$Zr \cite{Shus19}. Medical applications, such as the the Gadolinium Neutron Capture Therapy (GNCT) \cite{Nov13} and Boron Neutron Capture Therapy (BNCT) \cite{YF13} are also based on large neutron absorption by the nuclei. Other examples of neutron poisons are $^{153}$Cd with neutron absorption cross section of $2 \times 10^{4}$ barns and $^{135}$Xe with $3 \times 10^{6}$ barns, the largest known cross section for neutron induced reactions. $^{113}$Cd, a cadmium isotope  is often used in reactors as a neutron absorber-moderator.  The neutron absorption on $^{135}$Xe  has a cross section reaching atomic values, but for other neighboring isotopes the cross sections are inexplicably much smaller. A list of  empirical neutron capture cross sections on several nuclei is shown in Table  \ref{tab1}, with data extracted from Refs. \cite{Mughab2003,Sta07,BT14}.

No satisfactory explanation for the very large cross sections in some of the isotopes is available \cite{Cohen1971}. In Ref. \cite{CHK2016} it was suggested that the reaction proceeds via the population of a single 1p-2h doorway in the compound nucleus, connected with an intermediate structure \cite{BF1963,KRY1963,Feshbach1993}. But a detailed calculation is sill missing in the literature. The phenomenon is most likely of statistical nature, based on a fortuitous  isolated compound nucleus resonance. The problem with this idea is that such a resonance requires very narrow conditions on its position and strength, which are seldom met. The probability that the reaction hits this resonance is given by \cite{CHK2016}
\begin{equation}
P(\eta_0) = \frac{1}{2\pi}\frac{1}{1 + \eta_{0}},
\end{equation}
with $\eta_{0} \equiv \Gamma_{D, n}/\Gamma_{q, n}$ as a measure of the enhancement due to the doorway state, where $\Gamma_{D,n}$ is the doorway width and $\Gamma_{q,n}$ is the compound neutron width.  But, while $\Gamma_{D,n}$ is in the keV region, $\Gamma_{q,n}$ is in the eV region. Therefore,  $\eta_{0}  \gg 1$, $P(\eta_0) \ll 1$, and the probability for the occurrence of the doorway enhancement is very small. Since most of the neutron capture cross sections are  inhibited by statistical fluctuations, the probable cause for the large neutron poison cross sections  remain within the domain of random phenomena.

\begin{table}
\begin{center}
\begin{tabular}
[c]{|l|l|l|l|}\hline
 Nucleus & Cross section (barn)  \\  \hline
$^9{\rm Be}$ &[8.77$ \pm 0.35] \times \text  10^{-3}$  \\
$^{\rm10}{\rm B}$ & 0.5$\pm$0.0.1 \\
$^{14}{\rm N}$ & [79.8$\pm1.4]\times \text 10^{-3}$ \\
$^{15}{\rm N}$ & [0.024$\pm0.008]\times \text10^{-3}$ \\
$^{16}{\rm O}$ & [0.19$\pm0.019 ]\times \text 10^{-3}$\\
$^{20}{\rm Ne}$&[37$\pm4]\times \text10^{-3}$\\
$^{21}{\rm Ne}$&0.666$\pm0.110$\\
$^{28}{\rm Si}$ & [177$\pm 5]\times \text10^{-3}$ \\
$^{40}{\rm Ar}$ & 0.660$\pm$0.01 \\
$^{40}{\rm Ca}$ & 0.41$\pm$ 0.02 \\
$^{56}{\rm Fe}$ & 2.59 $\pm$0.14 \\
$^{59}{\rm Co}$ & 37.18 $\pm$ 0.06 \\
$^{58}{\rm Ni}$ & 4.5$\pm$0.2 \\
$^{63}{\rm Cu}$ & 4.52$\pm$ 0.02 \\  \hline
\end{tabular}
\begin{tabular}
[c]{|l|l|l|l|}\hline
 Nucleus & Cross section (barn)  \\  \hline
$^{84}{\rm Kr}$ & 0.111$\pm$0.015 \\
$^{88}{\rm Zr}$ & [8.61$\pm$0.69]$\times \text 10^{5}$ \\
$^{103}{\rm Rh}$ & 145$\pm$2 \\
$^{113}{\rm Cd}$ & [2.06$\pm0.04$]$\times \text 10^{4}$ \\
$^{114}{\rm Cd}$ & 0.34$\pm$0.02 \\
$^{135}${\rm Xe} & $2.65 \times 10^6$ \cite{Sta07}\\
$^{136}${\rm Xe} & $\sim 1 \times 10^{-3}$ \cite{BT14}\\
$^{149}{\rm Sm}$ & [4.014$\pm 0.06$]$\times \text 10^{4}$ \\
$^{157}{\rm Gd}$ & [2.54$\pm0.008$]$\times \text 10^{5}$ \\
$^{159}{\rm Tb}$ & 23.3$\pm$0.4 \\
$^{208}{\rm Pb}$ & [0.23$\pm0.03$]$\times 10^{-3}$ \\
$^{209}{\rm Bi}$ & 0.0338$\pm$0.0007 \\
$^{232}{\rm Th}$ & 7.35$\pm$0.03 \\
$^{238}{\rm U}$ & 2.68$\pm$ 0.019\\ \hline
\end{tabular}\caption{{Neutron absorption cross sections in several nuclei for epithermal neutrons ($E_n \lesssim 0.4$ eV). We have chosen adjacent nuclei across different mass regions to display the very different values of their  neutron absorption cross sections. The compiled data of experimental results were taken from Refs.  \cite{Mughab2003,Sta07,BT14,Shus19}.}}
\label{tab1}
\end{center}
\end{table}

In Ref. \cite{Bert19} the INEB formalism was further developed to relate the  neutron capture cross sections with (d,p) reactions in inverse kinematics.  It is readily noticed that the zero point motion of the neutron inside the deuteron has a large effect in reducing the neutron absorption cross section from the free neutron values by several orders of magnitudes. It is also found out that the best energies for such studies is about 30 MeV/nucleon ions incident on deuteron targets. In Table \ref{tb2} we show a numerical evaluation  for  the contribution of the $s$ and $d$ deuteron bound states to the neutron capture cross sections \cite{Bert19}. The purpose of these calculations is to motivate experiments with neutron poisons. In fact, for projectiles such as  $^{135}$Xe and $^{157}$Gd large cross sections are also obtained for deuteron targets. These cross sections are amenable to experimental studies using radioactive beams of  $^{135}$Xe and $^{157}$Gd.

\begin{table}
\begin{center}
\begin{tabular}
[c]{|l|l|l|l|}\hline
 Nucleus & $\sigma^s$ (mb) & $\sigma^d$ (mb)  \\  \hline
$^{59}{\rm Co}$ & $2.01 \times 10^{-2}$ & $5.08 \times 10^{-3}$   \\
$^{58}{\rm Ni}$ & $1.63\times 10^{-3}$& $9.26 \times 10^{-4}$ \\
$^{63}{\rm Cu}$ & $2.41 \times 10^{-3}$ & $9.06 \times 10^{-4}$\\
$^{88}{\rm Zr}$ & ${1.54\times 10^{2}}$ & ${0.37 \times 10^{2}}$ \\
$^{135}${\rm Xe} & ${7.02 \times 10^2}$  & ${2.85 \times 10^2}$ \\ \hline
\end{tabular}
\begin{tabular}
[c]{|l|l|l|l|}\hline
 Nucleus & $\sigma^s$ (mb)  & $\sigma^d$ (mb) \\  \hline
$^{149}{\rm Sm}$ & $11.3$ &3.48 \\
$^{157}{\rm Gd}$ & ${7.31 \times 10^1}$&${17.3}$ \\
$^{159}{\rm Tb}$ & $7.29\times 10^{-3}$&$1.55\times 10^{-3}$ \\
$^{232}{\rm Th}$ & $2.76\times 10^{-3}$&$8.07\times 10^{-3}$ \\
$^{238}{\rm U}$ & $4.61\times 10^{-4}$& $2.25 \times 10^{-4}$\\ \hline
\end{tabular}
\caption{{Neutron absorption cross section for (d,p) reactions  for several target nuclei. Effective deuteron energies of 30 MeV in inverse kinematics were considered. Note the large cross sections, within experimental reach, for $^{88}{\rm Zr}$, $^{135}$Xe and $^{157}$Gd.} }
\label{tb2}
\end{center}
\end{table}

\subsection{Statistical coupled-channels theory}
Numerical calculations for elastic and inelastic scattering of a system involving a very large number of channels is often a prohibitive computational task. Reactions with radioactive beams often involves this feature, and in particular, for weakly bound nuclei the continuum might need a special treatment with a continuum discretization, often known as continuum-discretized-coupled-channels (CDCC)  \cite{Kam86,Aus87,KY16}. For a comprehensive review of the CDCC method, see \cite{Mas212}. Practical continuum discretization methods have been discussed in Ref. \cite{BC92}  that can also be used in semiclassical calculations.

A few works have pursued the challenging task to implement a full numerical implementation of reaction cross sections for nucleon-nucleus scattering including the coupling of the elastic channel with a large number of particle-hole excitations in the nucleus. The particle-hole states are regarded as doorway states through which the reaction flows to more complicated nuclear configurations, followed by a flux to long-lived compound nucleus resonances. In Ref. \cite{Nob10} the many channels   were microscopically obtained from a random-phase calculation with a Skyrme force. A good agreement was obtained for proton and neutron scattering at 10-40 MeV incident energies. In Ref. \cite{BDH14} alternative statistical methods were proposed to mitigate the problem of coupling to an intractable large number of channels. Two relevant theoretical approaches have been developed based on the average of a large number of couplings to background channels  while keeping the couplings of main doorway channels fully in the coupled-channels procedure. But, despite the appealing feature of simplifying continuum-discretized-coupled-channels calculations, and of many doorway states, the statistical CDCC equations developed in Ref. \cite{BDH14} have not been implemented numerically. 

To complement the techniques developed in Ref. \cite{BDH14}, we introduce another method based on random matrices to treat coupled channels with a very large number of states.    
We will describe reactions with many couplings to break-up channels.  We take the CDCC Hamiltonian to be 
\begin{equation}
H=H_{0}+T_{R}+V_{tc}\left({\bf R}+\frac{A_{f}}{A_{p}}{\bf r}\right)+V_{tf}\left({\bf R}-\frac{A_{c}}{A_{p}}{\bf r}\right),
\end{equation}
where $H_0$ is the nuclear intrinsic Hamiltonian, $c$ and $f$ are the two fragments of the projectile. We expand
the partial wave components of the wave function as 
\begin{equation}
\Psi^{JM\pi}\left({\bf R},{\bf r}\right)=\frac{1}{rR}\sum_{ljLi}u_{ljLi}^{J\pi}(R)\phi_{li}^{j}(r)Y_{ljL}^{JM}\left(\Omega_{R,}\Omega_{r}\right),
\end{equation}
with 
\begin{equation}
Y_{ljL}^{JM}\left(\Omega_{R,}\Omega_{r}\right)=i^{l+L}\left[\left(Y_{l}\left(\Omega_{r}\right)\otimes\chi_{s}\right)^{j}\otimes Y_{L}\left(\Omega_{R}\right)\right]^{JM}.
\end{equation}
In the equations above, $\phi_{li}^{j}(r)$ are the wave functions describing the relative
motion of the projectile fragments $c$ and $f$, satisfying 
\begin{equation}
H_{0}\phi_{li}^{j}=\varepsilon_{li}^{j}\phi_{li}^{j},
\end{equation}
and the $u_{ljLi}^{J\pi}(R)$ the wave functions describing the relative
motion between the projectile center of mass and the target. These
satisfy the equation 
\begin{align}
  \left[-\frac{\hbar^{2}}{2\mu}\left(\frac{d^{2}\;}{dR^{2}}-\frac{L(L+1)}{R^{2}}\right)+\right. & \left. U_{c}(R)+\varepsilon_{c}-E\right]u_{c}^{J\pi}(R) \nonumber\\
   = & \sum_{c^{\prime}}v_{cc^{\prime}}(R)u_{c^{\prime}}^{J\pi}(R),
\end{align}
where we have now substituted the index c for the set ($ljLi$), with
the understanding that $\varepsilon_{c}=\varepsilon_{li}^{j}$. We
have divided the interaction into a diagonal part $U_{c}(R)$, which
contains the monopole nuclear and Coulomb potentials, and a term $v_{cc^{\prime}}(R)$,
which describes the coupling among the channels.

To solve these equations we expand the wave function in the internal
region ($0\le R\le a$) in a basis $\varphi_{n}(R)$, as 
\begin{equation}
u_{c,int}^{J\pi}(R)=\sum_{n}A_{cn}^{J\pi}\varphi_{n}(R),\qquad R\le a,
\end{equation}
and match at $R=a$ to the Coulomb wave functions of the external
wave function, 
\begin{equation}
u_{c,ext}^{J\pi}(R)=\frac{1}{\sqrt{v_{C}}}\left(I_{c}\left(k_{c}R\right)\delta_{cc_{0}}-O_{c}\left(k_{c}R\right)U_{cc_{0}}^{J\pi}\right),
\end{equation}
where $v_{c}$ and $k_{c}$ are the relative velocity and wave number
in the channel, $O_{c}=I_{c}^{*}$ are the outgoing and incoming Coulomb
waves and $U_{cc^{\prime}}^{J\pi}$is the scattering matrix.\cite{PD2010}

We use the Bloch operator \cite{BF1963,KRY1963,Feshbach1993}
\begin{equation}
{\mathcal{L}}_{c}=\frac{\hbar^{2}}{2\mu}\delta(R-a)\left(\frac{d\;}{dR}-B_{c}\right),
\end{equation}
where we will take $B_{c}=0$. Then we define a matrix
$C$ as 
\begin{align}
  C_{cn,c^{\prime}n^{\prime}}^{J\pi} & \\
  =\left\langle\varphi_{n}\right. & \left.\left|\left(T_{c}+U_{c}+\varepsilon_{c}+{\mathcal{L}}_{c}-E\right)\delta_{cc^{\prime}}+V_{cc^{\prime}}^{J\pi}\right|\varphi_{n^{\prime}}\right\rangle \nonumber
\end{align}
and the R-matrix as
\begin{equation}
R_{cc^{\prime}}^{J\pi}=\sum_{dnd^{\prime}n^{\prime}}\gamma_{c,dn}\left[\frac{1}{C^{J\pi}}\right]_{dn,d^{\prime}n^{\prime}}\gamma_{c^{\prime},d^{\prime}n^{\prime}},
\end{equation}
where 
\begin{equation}
\gamma_{c,dn}=\sqrt{\frac{\hbar^{2}}{2\mu a}}\varphi_{n}(a)\delta_{cd}.
\end{equation}
Suppressing the indices, we can write 
\begin{equation}
R^{J\pi}=\gamma\frac{1}{C^{J\pi}}\gamma^{T}.
\end{equation}
We can now write the scattering matrix $U^{J\pi}$in terms of $R^{J\pi}$as
\begin{equation}
U^{J\pi}=\rho^{1/2}O\left(1-R^{J\pi}D\right)^{-1}\left(1-R^{J\pi}D^{*}\right)I\rho^{-1/2},
\end{equation}
where 
\begin{eqnarray}
\rho_{cc^{\prime}} & = & k_{c}a\delta_{cc^{\prime}}, \nonumber\\
O_{cc^{\prime}} & = & O_{c}(k_{c}a)\delta_{cc^{\prime}},\\
I_{cc^{\prime}} & = & I_{c}(k_{c}a)\delta_{cc^{\prime}} \nonumber
\end{eqnarray}
and 
\begin{equation}
D_{cc^{\prime}}=k_{c}a\frac{O_{c}^{\prime}(k_{c}a)}{O_{c}(k_{c}a)}\delta_{cc^{\prime}}=k_{c}a\left[\frac{I_{c}^{\prime}(k_{c}a)}{I_{c}(k_{c}a)}\right]^{*}\delta_{cc^{\prime}}.
\end{equation}

We now write 
\begin{eqnarray}
  \left(1-R^{J\pi}D\right)^{-1} & = & 1+R^{J\pi}D \nonumber\\
  & & \qquad +R^{J\pi}DR^{J\pi}D+\dots\nonumber\\
  & = & 1+\gamma\frac{1}{C^{J\pi}}\gamma^{T}D \\
  & & \; +\gamma\frac{1}{C^{J\pi}}\gamma^{T}D\gamma\frac{1}{C^{J\pi}}\gamma^{T}D+\dots\nonumber \\
 & = & 1+\gamma\frac{1}{C^{J\pi}-\gamma^{T}D\gamma}\gamma^{T}D.\nonumber
\end{eqnarray}
Substituting, we may reduce 
\begin{align}
  \left(1-R^{J\pi}D\right)^{-1} & \left(1-R^{J\pi}D^{*}\right)\\
  = & 1+\gamma\frac{1}{C^{J\pi}-\gamma^{T}D\gamma}\gamma^{T}\left(D-D^{*}\right)\nonumber
\end{align}
and write the scattering matrix as 
\begin{equation}
U^{J\pi}=\rho^{1/2}O\left(1+\gamma\frac{1}{C^{J\pi}-\gamma^{T}D\gamma}\gamma^{T}\left(D-D^{*}\right)\right)I\rho^{-1/2}.
\end{equation}
Now let us assume that we are only interested in a strongly-coupled
subset of the states $c$. We can the write the matrix as 
\begin{equation}
C^{J\pi}-\gamma^{T}D\gamma=\left(\begin{array}{cc}
C_{0}^{J\pi}-\gamma^{T}D_{0}\gamma & V\\
V^{\dagger} & C_{x}^{J\pi}-\gamma^{T}D_{x}\gamma
\end{array}\right),
\end{equation}
where the sub-matrix containing the set of interest is denoted by $0$
and the remaining set by $x$. Since the matrix elements of $\gamma^{T}D\gamma$
are diagonal in $c$ (but not in $n$), the only matrix elements coupling
the two sets of states are the (weak) interaction terms $v_{cn,c^{\prime}n^{\prime}}$,
which we denote here by $V.$ Its inverse is given by {\footnotesize{{{{
\begin{widetext}
\begin{equation}
\begin{array}{c}
\left(\begin{array}{cc}
\left(C_{0}^{J\pi}-\gamma^{T}D_{0}\gamma-V\frac{1}{C_{x}^{J\pi}-\gamma^{T}D_{x}\gamma}V^{\dagger}\right)^{-1} & \left(C_{0}^{J\pi}-\gamma^{T}D_{0}\gamma-V\frac{1}{C_{x}^{J\pi}-\gamma^{T}D_{x}\gamma}V^{\dagger}\right)^{-1}V\frac{1}{C_{x}^{J\pi}-\gamma^{T}D_{x}\gamma}\\
\frac{1}{C_{x}^{J\pi}-\gamma^{T}D_{x}\gamma}V^{\dagger}\left(C_{0}^{J\pi}-\gamma^{T}D_{0}\gamma-V\frac{1}{C_{x}^{J\pi}-\gamma^{T}D_{x}\gamma}V^{\dagger}\right)^{-1} & \left(C_{x}^{J\pi}-\gamma^{T}D_{x}\gamma-V\frac{1}{C_{0}^{J\pi}-\gamma^{T}D_{0}\gamma}V^{\dagger}\right)^{-1}
\end{array}\right)
\end{array}
\end{equation}
\end{widetext}
}}}}}{\footnotesize \par}

Up to this point, the development has been exact. To simplify the
calculation of the component of interest, we would like to approximate
the matrix 
\begin{equation}
V\frac{1}{C_{x}^{J\pi}-\gamma^{T}D_{x}\gamma}V^{\dagger}\,.
\end{equation}
We would also like to approximate the term 
\begin{equation}
V\frac{1}{C_{x}^{J\pi}-\gamma^{T}D_{x}^{*}\gamma}\gamma^{T}\,\frac{\pi}{k_{x}^{2}}\rho_{x}\left|O_{x}\right|^{2}\,\gamma\frac{1}{C_{x}^{J\pi}-\gamma^{T}D_{x}\gamma}V^{\dagger}\,,
\end{equation}
which enters the expression for the summed cross section of the weakly-coupled
states.

\subsubsection*{Weak DWBA }
If the coupling among the channels weakly-coupled to the entrance
channel is also weak, the simplest approximation would be to neglect
the coupling term in $C_{x}^{J\pi}$, a type of DWBA approximation.
In that case, we would take 
\begin{align}
  \left(C_{x}^{J\pi}\right. & \left.\right)_{cn,c^{\prime}n^{\prime}} 
  \\\rightarrow & \left\langle \varphi_{n}\left|T_{c}+U_{c}+\varepsilon_{c}+{\mathcal{L}}_{c}-E\right|\varphi_{n^{\prime}}\right\rangle \delta_{cc^{\prime}}\nonumber
\end{align}
and 
\begin{align}
  \left(C_{x}^{J\pi}\right. & \left.-\gamma^{T}D_{x}\gamma\right)_{cn,c^{\prime}n^{\prime}}  \\
  \rightarrow & \left(\left\langle \varphi_{n}\left|T_{c}+U_{c}+\varepsilon_{c}+{\mathcal{L}}_{c}-E\right|\varphi_{n^{\prime}}\right\rangle -\frac{\hbar^{2}}{2\mu a}D_{c}\right)\delta_{cc^{\prime}}\,.\nonumber
\end{align}
We would then have 
\begin{align}
  \Big(V\Big.&\Big.\frac{1}{C_{x}^{J\pi}-\gamma^{T}D_{x}\gamma}V^{\dagger}\Big)_{c_{0}n,c_{0}^{\prime}n^{\prime}}  \\
  \rightarrow & \sum_{c_{x}mm^{\prime}}v_{c_{0}n,c_{x}m}\left(\frac{1}{C_{x}^{J\pi}-\gamma^{T}D_{x}\gamma}\right)_{c_{x}m,c_{x}m^{\prime}}v_{c_{x}m^{\prime},c_{0}^{\prime}n^{\prime}}\,.\nonumber
\end{align}

The coupling through the weakly-coupled terms would then include only
their optical propagation part - the $c_{x}$ submatrices would not
be coupled. The R-matrix solution would still require solution of
each $c_{x}$ channel in the expansion set $\left\{ \varphi_{n}\right\} $
but would not require calculation and diagonalization of the $c_{x}n$
matrix.

\subsubsection*{Statistical CDCC}

Alternatively, if the coupling among the channels weakly-coupled to
the entrance channel is sufficiently strong to mix them, we can consider
using statistical hypotheses in approximations to the cross sections.
If the expansion functions $\phi_{li}^{j}(r)$ and $\varphi_{n}(R)$
are real, the matrix $C_{x}^{J\pi}$ is real symmetric and the matrix
$C_{x}^{J\pi}-\gamma^{T}D_{x}\gamma$ complex symmetric. There is
thus a unitary transformation $U$ for which $U\left(C_{x}^{J\pi}-\gamma^{T}D_{x}\gamma\right)U^{T}$
is diagonal. This does not look to be that useful an expression, since
we cannot expect special behavior of the matrix elements of $UU^{T}$.

On the other hand, whether the basis functions are real or not, the
fact that $C_{x}^{J\pi}$ is Hermitian implies that there exist a
$U$ for which $UC_{x}^{J\pi}U^{\dagger}$ is diagonal. In this case
we have 
\begin{equation}
V\frac{1}{C_{x}^{J\pi}}V^{\dagger}=VU\frac{1}{{\mathcal{E}}_{x}-E}U^{\dagger}V^{\dagger},
\end{equation}
where ${\mathcal{E}}_{x}$ is the diagonalized Hamiltonian in the
$x$ subspace. Rewriting the matrix elements $VU$ as 
\begin{equation}
\left(VU\right)_{c_{0}n,\lambda}=\sum_{c_{x}^{\prime}n^{\prime}}v_{c_{0}n,c_{x}^{\prime}n^{\prime}}U_{c_{x}^{\prime}n^{\prime},\lambda},
\end{equation}
we expect that we might find 
\begin{align}
  \Big(VU\frac{1}{{\mathcal{E}}_{x}-E}\Big.&\Big.U^{\dagger}V^{\dagger}\Big)_{c_{0}n,c_{0}^{\prime}n^{\prime}} \\
  = & \sum_{\lambda}\left(VU\right)_{c_{0}n,\lambda}\frac{1}{\epsilon_{\lambda}-E}\left(U^{\dagger}V^{\dagger}\right)_{\lambda,c_{0}^{\prime}n^{\prime}}\nonumber\\
  \approx &  \sum_{\lambda}\left|\left(VU\right)_{c_{0}n,\lambda}\right|^{2}\frac{1}{\epsilon_{\lambda}-E}\delta_{c_{0}c_{0}^{\prime}}\delta_{nn^{\prime}}\nonumber
\end{align}
due to the fact that matrix elements with different values of $c$
and $n$ will have different phases and will tend to average to zero.

To perform calculations, a stronger version of this hypothesis is
needed. First, we require the number of states $N_{\Lambda}$ to be large in intervals
$\Delta\epsilon_{\lambda}$ in which optical quantities, such as $k$,
$\rho,$ and $D$ vary slowly and denote these intervals by $\Lambda.$
We then assume that 
\begin{equation}
\sum_{\lambda\in\Lambda}\left(VU\right)_{c_{0}n,\lambda}\left(U^{\dagger}V^{\dagger}\right)_{\lambda,c_{0}^{\prime}n^{\prime}}=N_{\Lambda}\overline{\left|\left(VU\right)_{c_{0}n,\Lambda}\right|^{2}}\delta_{c_{0}c_{0}^{\prime}}\delta_{nn^{\prime}}\,,
\end{equation}
so that 
\begin{align}
  \Big(VU\frac{1}{{\mathcal{E}}_{x}-E}U^{\dagger}\Big.&\Big.V^{\dagger}\Big)_{c_{0}n,c_{0}^{\prime}n^{\prime}} \\
  & =\sum_{\lambda\in\Lambda}\overline{\left|\left(VU\right)_{c_{0}n,\Lambda}\right|^{2}}\frac{N_{\Lambda}}{\epsilon_{\Lambda}-E}\delta_{c_{0}c_{0}^{\prime}}\delta_{nn^{\prime}}\,.\nonumber
\end{align}
However, the matrix element we want to simplify is 
\begin{align}
  \Big(V\Big.&\Big.\frac{1}{C_{x}^{J\pi}-\gamma^{T}D_{x}\gamma}V^{\dagger}\Big)_{c_{0}n,c_{0}^{\prime}n^{\prime}}\\
  &\qquad =\left(VU\frac{1}{{\mathcal{E}}_{x}-E-U^{\dagger}\gamma^{T}D_{x}\gamma U}U^{\dagger}V^{\dagger}\right)_{c_{0}n,c_{0}^{\prime}n^{\prime}}\,.\nonumber
\end{align}
This will be easy to calculate if 
\begin{eqnarray}
  \left(U^{\dagger}\gamma^{T}D_{x}\gamma U\right)_{\lambda\lambda\prime} & = & \sum_{c_{x}nn^{\prime}}\left(U^{\dagger}\gamma^{T}\right)_{\lambda,c_{x}n}D_{c_{x}}\left(\gamma U\right)_{c_{x}n^{\prime}\lambda^{\prime}}\nonumber\\
  &\rightarrow & \left(\overline{\gamma^{T}D_{x}\gamma}\right)_{\Lambda}\delta_{\lambda\lambda^{\prime}}\\
&  &\equiv  i\Gamma_{\Lambda}/2\,\delta_{\lambda\lambda^{\prime}}\,,\nonumber
\end{eqnarray}
for $\lambda\in\Lambda$, which we will assume to be the case. We then have 
\begin{align}
  \Big(V\Big.&\Big.\frac{1}{C_{x}^{J\pi}-\gamma^{T}D_{x}\gamma}V^{\dagger}\Big)_{c_{0}n,c_{0}^{\prime}n^{\prime}}\\
  & =  \left(VU\frac{1}{{\mathcal{E}}_{x}-i\Gamma_{x}/2-E}U^{\dagger}V^{\dagger}\right)_{c_{0}n,c_{0}^{\prime}n^{\prime}} \nonumber\\
 & \quad\rightarrow  \sum_{\Lambda}\overline{\left|\left(VU\right)_{c_{0}n,\Lambda}\right|^{2}}\frac{N_{\Lambda}}{\epsilon_{\Lambda}-i\Gamma_{\Lambda}/2-E}\delta_{c_{0}c_{0}^{\prime}}\delta_{nn^{\prime}}.\nonumber 
\end{align}
If the average matrix elements vary smoothly with $\Lambda$, we can
write ($N_{\Lambda}\rightarrow \rho_{x}\left(\epsilon_{\Lambda}\right)d\epsilon_{\Lambda}$)
\begin{align}
  \sum_{\Lambda}\overline{\left|\left(VU\right)_{c_{0}n,\Lambda}\right|^{2}}&\frac{N_{\Lambda}}{\epsilon_{\Lambda}-i\Gamma_{\Lambda}/2-E} \\
  & =  \int\overline{\left|\left(VU\right)_{c_{0}n,\Lambda}\right|^{2}}\frac{\rho_{x}\left(\epsilon_{\Lambda}\right)d\epsilon_{\Lambda}}{\epsilon_{\Lambda}-i\Gamma_{\Lambda}/2-E}\nonumber\\
 & \equiv  \Delta_{c_{0}n}(E)+i\Gamma_{c_{0}n}^{\downarrow}\left(E\right)/2 \nonumber\\
 & \approx  2\pi i\overline{\left|\left(VU\right)_{c_{0}n,\Lambda}\right|^{2}}\rho_{x}\left(\epsilon_{\Lambda}=E\right).\nonumber
\end{align}
If the statistical hypotheses are consistent with the physics of the
problem, this is equivalent to
\begin{align}
  \Big(V\Big.&\Big.\frac{1}{C_{x}^{J\pi}-\gamma^{T}D_{x}\gamma}V^{\dagger}\Big)_{c_{0}n,c_{0}^{\prime}n^{\prime}}\\
  & \qquad\qquad =\left(\Delta_{c_{0}n}(E)+i\Gamma_{c_{0}n}^{\downarrow}\left(E\right)/2\right)\delta_{c_{0}c_{0}^{\prime}}\delta_{nn^{\prime}}\,.\nonumber
\end{align}
 We can then approximate the S-matrix of the strongly-coupled states
as 
\begin{align}
  U_{0}^{J\pi} & \\
   & =\rho_{0}^{1/2}O_{0}\bigg(1 +\gamma\frac{1}{C_{0}^{J\pi}-\gamma^{T}D_{0}\gamma-\Delta_{0}-i\Gamma_{0}^{\downarrow}/2}\bigg.\nonumber\\
  & \qquad\qquad\qquad\qquad\qquad\bigg.\times\gamma^{T}\left(D_{0}-D_{0}^{*}\right)\bigg)I_{0}\rho_{0}^{-1/2}\nonumber
\end{align}
and can calculate the cross sections accordingly. Note that these
cross sections only take into account the loss of flux to the weakly-coupled
states. They do not include the fluctuation contribution due to coupling
through the weakly-coupling states back to the strongly-coupled ones.
The lowest order fluctuation contribution will come from the term
\begin{eqnarray}
  U_{0,fl}^{J\pi} & =& \rho_{x}^{1/2}O_{x}\gamma\frac{1}{C_{0}^{J\pi}-\gamma^{T}D_{0}\gamma-\Delta_{0}-i\Gamma_{0}^{\downarrow}/2}\\
& &  \qquad\times V\frac{1}{C_{x}^{J\pi}-\gamma^{T}D_{x}\gamma}V^{\dagger} \nonumber\\
  & & \qquad\times \frac{1}{C_{0}^{J\pi}-\gamma^{T}D_{0}\gamma-\Delta_{0}-i\Gamma_{0}^{\downarrow}/2}\nonumber\\
  & & \qquad\qquad\times\gamma^{T}\left(D_{0}-D_{0}^{*}\right)I_{0}\rho_{0}^{-1/2}\,,\nonumber
\end{eqnarray}
where the couplings $V$ and $V^{\dagger}$ must be averaged with those
of the conjugate contribution, $U_{0,fl}^{J\pi\dagger}$. The fluctuation
contribution is thus of fourth-order in the coupling $V$.

The lowest order contribution to the S-matrix elements of the channels
weakly coupled to the entrance channel are given by 
\begin{eqnarray}
  U_{x0}^{J\pi} & = & \rho_{x}^{1/2}O_{x}\gamma\frac{1}{C_{x}^{J\pi}-\gamma^{T}D_{x}\gamma}V^{\dagger}\\
  & & \qquad\times\frac{1}{C_{0}^{J\pi}-\gamma^{T}D_{0}\gamma-\Delta_{0}-i\Gamma_{0}^{\downarrow}/2}\nonumber\\
  & & \qquad\qquad\times\gamma^{T}\left(D_{0}-D_{0}^{*}\right)I_{0}\rho_{0}^{-1/2}\,,\nonumber
\end{eqnarray}
with cross sections given by 
\begin{equation}
\sigma_{x0}=\frac{\pi}{k_{x}}\left|U_{x0}^{J\pi}\right|^{2}\,.
\end{equation}
The inclusive cross section to the weakly-coupled states can be written
as 
\begin{eqnarray}
  \sum_{x}\sigma_{x0} & = & S^{\dagger}V\frac{1}{C_{x}^{J\pi}-\gamma^{T}D_{x}^{*}\gamma}\gamma^{T}\,\frac{\pi}{k_{x}^{2}}\rho_{x}\left|O_{x}\right|^{2}\\
  & & \qquad\qquad\times\gamma\frac{1}{C_{x}^{J\pi}-\gamma^{T}D_{x}\gamma}V^{\dagger}S\,,\nonumber
\end{eqnarray}
where 
\begin{equation}
S=\frac{1}{C_{0}^{J\pi}-\gamma^{T}D_{0}\gamma-\Delta_{0}-i\Gamma_{0}^{\downarrow}/2}\gamma^{T}\left(D_{0}-D_{0}^{*}\right)I_{0}\rho_{0}^{-1/2}\,.
\end{equation}
We write 
\begin{align}
  \bigg(V\bigg.&\bigg.\frac{1}{C_{x}^{J\pi}-\gamma^{T}D_{x}^{*}\gamma}\gamma^{T}\,\frac{\pi}{k_{x}^{2}}\rho_{x}\left|O_{x}\right|^{2}\,\gamma\frac{1}{C_{x}^{J\pi}-\gamma^{T}D_{x}\gamma}V^{\dagger}\bigg)_{c_{0}n,c_{0}^{\prime}n^{\prime}}\nonumber\\
  & =  \bigg(VU\frac{1}{{\mathcal{E}}_{x}+i\Gamma_{x}/2-E}U^{\dagger}\gamma^{T}\,\frac{\pi}{k_{x}^{2}}\rho_{x}\left|O_{x}\right|^{2}\bigg.\nonumber\\
  & \qquad\qquad\qquad\qquad\bigg.\times\gamma U\frac{1}{{\mathcal{E}}_{x}-i\Gamma_{x}/2-E}U^{\dagger}V^{\dagger}\bigg)_{c_{0}n,c_{0}^{\prime}n^{\prime}}\nonumber\\
 & =  \sum_{\Lambda}\overline{\left|\left(VU\right)_{c_{0}n,\Lambda}\right|^{2}}\frac{N_{\Lambda}\Gamma_{\Lambda}^{\uparrow}/2}{\left|\epsilon_{\Lambda}-i\Gamma_{\Lambda}/2-E\right|^{2}}\delta_{c_{0}c_{0}^{\prime}}\delta_{nn^{\prime}}\,
\end{align}
where we have assumed that 
\begin{eqnarray}
\left(U^{\dagger}\gamma^{T}\,\frac{\pi}{k_{x}^{2}}\rho_{x}\left|O_{x}\right|^{2}\,\gamma U\right)_{\lambda\lambda^{\prime}} & = & \left(\overline{\gamma^{T}\,\pi\frac{\rho_{x}}{k_{x}^{2}}\left|O_{x}\right|^{2}\,\gamma}\right)_{\Lambda}\delta_{\lambda\lambda^{\prime}} \nonumber\\
 & \equiv & \Gamma_{\Lambda}^{\uparrow}/2\,\delta_{\lambda\lambda^{\prime}}.
\end{eqnarray}
Transforming the sum over states to an integral, as before, we find
\begin{align}
  \sum_{\Lambda}&\overline{\left|\left(VU\right)_{c_{0}n,\Lambda}\right|^{2}}\frac{N_{\Lambda}\Gamma_{\Lambda}^{\uparrow}/2}{\left|\epsilon_{\Lambda}-i\Gamma_{\Lambda}/2-E\right|^{2}} \nonumber \\
 & \rightarrow  \int\overline{\left|\left(VU\right)_{c_{0}n,\Lambda}\right|^{2}}\frac{\rho_{x}\left(\epsilon_{\Lambda}\right)\Gamma_{\Lambda}^{\uparrow}/2}{\left|\epsilon_{\Lambda}-i\Gamma_{\Lambda}/2-E\right|^{2}}\, d\epsilon_{\Lambda} \\
 & \qquad\approx  \pi\frac{\Gamma_{\Lambda}^{\uparrow}}{\Gamma_{\Lambda}}\rho_{x}\left(\epsilon_{\Lambda}=E\right)\overline{\left|\left(VU\right)_{c_{0}n,\Lambda}\right|^{2}}.\nonumber
\end{align}
The inclusive cross section to the channels weakly coupled to the
entrance channel can then be written as 
\begin{equation}
\sum_{x}\sigma_{x0}=\pi\frac{\Gamma_{\Lambda}^{\uparrow}}{\Gamma_{\Lambda}}\rho_{x}\left(\epsilon_{\Lambda}=E\right)\sum_{c_{0}^{\prime}n^{\prime}n}\overline{\left|\left(VU\right)_{c_{0}n,\Lambda}\right|^{2}}\left|S_{c_{0}^{\prime}n^{\prime},c_{0}n}\right|^{2}\,.
\end{equation}
These cross sections also have higher-order corrections corresponding
to coupling through the weakly-coupled states to the strongly-coupled
ones and then back to the weakly-coupled ones.

We note that the loss terms also have higher-order corrections coming
from averages over different pairs of interactions, such as the following
one, in which the middle two instances and the outer two instances
of the interaction are averaged 
\begin{equation}
\overline{V\frac{1}{C_{x}^{J\pi}-\gamma^{T}D_{x}\gamma}\overline{V^{\dagger}\frac{1}{C_{0}^{J\pi}-\gamma^{T}D_{0}\gamma}V}\frac{1}{C_{x}^{J\pi}-\gamma^{T}D_{x}\gamma}V^{\dagger}}\,.
\end{equation}
When the coupling is weak, this term can usually be neglected in comparison
to the term 
\begin{equation}
\overline{V\frac{1}{C_{x}^{J\pi}-\gamma^{T}D_{x}\gamma}V^{\dagger}}\,,
\end{equation}
for which it would serve as a correction in the development above.
(The average over the first pair of $V$'s and the second pair of
$V$'s furnishes the second-order iteration of the latter term in
the expansion rather than a correction to it.) When the number of
weakly-coupled states is large, it can be shown that the other possible
combination of averages - the first and third instances and the second
and fourth instances of $V$ - is much smaller than the other two. This method has not been implemented numerically, but work in this direction is in progress. It will simplify calculations of reaction cross sections involving  too many states, beyond the reach of present computing capabilities.

\section{Inclusive Non-Elastic Breakup Reactions of  two-fragment projectile on a two-fragment target}
We now consider a four-body breakup problem for  the reaction of a two-fragment projectile with a two-fragment target. Both projectile and target can dissociate into their two clusters, forming a genuine four-body system. The formalism of Ref. \cite{Hus20} can be applied by describing the target as a = b + $x_2$ cluster, and the projectile as another A = $x_1$ + B cluster. Therefore, the inclusive measurement of $b$ will include the breakup of the projectile with $x_1$ reacting with the target $a = x_1 + d$, and also the breakup of the target with $x_2$ reacting with the projectile, formed by the $x_2$ + A cluster. This reaction is a complicated high-energy four-body problem. To simplify it to a manageable problem we treat the reaction as a two three-body process. The breakup of the projectile is assumed to proceed without affecting the target and the breakup of the target is also assumed to proceed without affecting the projectile. One then obtains two separate groups of detected spectator fragments, one belonging to the target and the other belonging to the projectile. This approach can become valuable to treat reactions involving a  neutron or proton-rich projectile with targets such as the deuteron. 

Let us consider the reaction $^8$B + d, yielding  p + (n + $^8$B) $\rightarrow$ p + $^9$B, or  p + ($^7$Be + d). Note that $^8$B is a one proton halo with proton separation energy of  0.137 MeV. The reaction  p + (n + $^8$B)  leads to a neutron capture by a one-proton halo nucleus, whereas the reaction p + ($^7$Be + d) yields an incomplete fusion of the core of the halo nucleus with the deuteron target. Denoting the proton originating from the radioactive projectile by $p_1$ and that from the deuteron target breakup by $p_2$, we can write the inclusive non-elastic proton spectrum as
\begin{eqnarray}
  \frac{d^{2}\sigma_{p}}{dE_{p}d\Omega_{p}} & = & \rho(E_{p_2}) \hat{\sigma}_{R}({\rm n} + 8{\rm B}) \label{B} \\
  & & \qquad + \rho(E_{p_1}) \hat{\sigma}_{R}({\rm d} + 7{\rm Be}) + \cdots\nonumber
\end{eqnarray}
The first term on the right-hand side contains the neutron capture cross section of the halo nucleus and is peaked at the higher proton energy  since the proton separation energy in the deuteron is 2.22 MeV.  In contrast, the second term corresponds to the incomplete fusion in the form of $^7$Be + d which leads to the emission of the halo proton in $^8$B first, followed by the collision of its core, $^7$Be, with the deuteron. The last reaction mechanism will dominate the low energy part of the inclusive proton spectrum. 

For a one-neutron halo projectile such as $^{11}$Be or $^{19}$C,
the same type of reaction yields an inclusive proton spectrum that will exhibit a low energy peak associated with the deuteron breakup at 2.22 MeV, and a weaker higher energy  peak related to the removal of a proton from the tightly bound cores, $^{10}$Be, $^{18}$C. That is,
\begin{eqnarray}
  \frac{d^{2}\sigma_{p}}{dE_{p}d\Omega_{p}} & = & \rho(E_{p_2}) \hat{\sigma}_{R}({\rm n} + ^{11}{\rm Be})\label{Be}\\
  & & \qquad + \rho(E_{p_{1}}) \hat{\sigma}_{R}({\rm d} +  ^{10}{\rm Be}) + \cdots\nonumber
\end{eqnarray}
\begin{eqnarray}
  \frac{d^{2}\sigma_{p}}{dE_{p}d\Omega_{p}} & = & \rho(E_{p_2}) \hat{\sigma}_{R}({\rm n} + ^{19}{\rm C})\label{C}\\
  & & \qquad + \rho(E_{p_{1}}) \hat{\sigma}_{R}({\rm d} + ^{18}{\rm B}) + \cdots\nonumber
\end{eqnarray}
The cross sections, $\hat{\sigma}_{R}(n + ^{11}$Be), $\hat{\sigma}_{R}(n + ^{19}$C), $\hat{\sigma}_{R}(d + ^{10}$Be), $\hat{\sigma}_{R}(d + ^{18}$B), are obtained from expressions similar to Eq. (\ref{sigRpar}). One needs the S-matrix elements,  
$\hat{S}_{p_{1}}(\textbf{r}_{p_{1}})$ and $\hat{S}_{p_{2}}(\textbf{r}_{p_{2}})$ to evaluate the  cross sections. The matrix elements can be calculated once the optical potentials appropriate for the scattering of protons on deuterons and on the different halo projectiles are known. Optical potentials for the projectile and target systems are also needed, as well as those to obtain  distorted waves  of the participant fragments. These are $n + ^{11}$Be, $n + ^{19}$C, $d + ^{10}$Be, $d + ^{18}$B.
In the case of the proton halo nucleus $^8$B, we need similar ingredients: $\hat{S}_{p_{1}}(\textbf{r}_{p_{1}})$ for p + d elastic scattering and  $\hat{S}_{p_{2}}(\textbf{r}_{p_{2}})$ for p + $^8$B. Similarly one needs the optical potentials for d + $^8$B,  n + $^8$B and 
d + $^7$Be. These potentials could be extracted for a phenomenological fit to elastic scattering data.

Once the incomplete fusion cross sections are calculated from fusion theory \cite{CGDH06}, the neutron capture cross sections can be obtained from the general form of the breakup cross sections, Eqs. (\ref{B}, \ref{Be}, \ref{C}).
Thus the Inclusive Non-Elastic Breakup is a potentially powerful method to extract the neutron capture cross section of short-lived radioactive nuclei.

\section{Inclusive Non-Elastic Breakup Reactions of three cluster projectiles with two cluster targets}
Let us consider three-cluster projectiles, $a = b + x_1 +x_2$, such as $^9$Be = $^4$He + $^4$He + n and Borromean nuclei such as $^{11}$Li = $^9$Li + n + n. The cross section for the four-body process, $ b + x_1 + x_2 + A$, where $b$ is the observed spectator fragment and $x_1$ and $x_2$ are the interacting participants fragments, is given by \cite{Hussein1984}
\begin{equation}
\frac{d^{2}\sigma^{INEB}_b}{dE_{b}d\Omega_{b}} = \rho_{b}(E_b)\sigma_{R}^{4B},
\end{equation}
\begin{equation}
\sigma_{R}^{4B} = \frac{k_a}{E_a}\left[\frac{{E_{x_1}}}{{k_{x_1}}}\sigma_{R}^{x_1} + \frac{{E_{x_2}}}{{k_{x_2}}}\sigma_{R}^{x_2} + \frac{E_{CM}({{x}_1},{{ x}_2})}{(k_{{x}_1}+ k_{{x}_2})} \sigma_{R}^{3B}\right],\label{CFH-R}
\end{equation}
where, 
\begin{eqnarray}
  \sigma_{R}^{x_1} & = &  \frac{{k_{x_1}}}{E_{{x_1}}} \langle \hat{\rho}_{{x}_1, {x}_2}|W_{x_1}|\hat{\rho}_{{x}_1, {x}_2}\rangle,\\
\sigma_{R}^{x_2} & = & \frac{{k_{x_2}}}{E_{{x_2}}} \langle \hat{\rho}_{{x}_1, {x}_2}|W_{x_2}|\hat{\rho}_{{x}_1, {x}_2}\rangle, \label{sigma_2}
\end{eqnarray}
and,
\begin{equation}
\sigma_{R}^{3B} = \frac{(k_{{x}_1}+ k_{{x}_2})}{E_{CM}({{x}_1},{{ x}_2})}\ \langle \hat{\rho}_{{x}_1, {x}_2}|W_{3B}|\hat{\rho}_{{x}_1, {x}_2}\rangle ,\label{sigma_3B}
\end{equation}
is a three-body, $x_1 + x_2 + A$, reaction cross section. The different energies of the fragments are determined by the beam energy for weakly bound projectiles with the binding energy being marginally relevant for the energies of the three outgoing fragments. In this case, we can use  e.g., $E_{{x_1},Lab} = E_{a, Lab}(M_{{x_{1}}}/M_a)$, with $M_{a}$ and $M_{{x_1}}$ being the mass numbers of the projectile and of the  fragment $x_1$, respectively. The three-body source function, $\hat{\rho}_{{x}_1, {x}_2}$, is a generalization of the two-body source function defined in Eqs. (\ref{rho1},\ref{rho2}), i.e.,
\begin{align}
\hat{\rho}_{x_1, x_2} (\textbf{r}_{x_{1}}, & \textbf{r}_{x_{2}})  \\
= & (\chi^{(-)}_{b} (\textbf{r}_b)|  \chi^{(+)}_{a}(\textbf{r}_{b}, \textbf{r}_{x_{1}}, \textbf{r}_{x_{2}})\Phi_{a}(\textbf{r}_{b}, \textbf{r}_{x_{1}}, \textbf{r}_{x_{2}})\rangle .\nonumber
\end{align}

The cross sections $\sigma_{R}^{x_1}$ and $\sigma_{R}^{x_2}$ are the reaction cross sections of $x_1$ + A and of $x_2$ + A whereas the other clusters, $x_2$ and $x_1$  are scattered and not observed.  In analogy with the developments presented in the previous section, one gets \cite{Hussein1984}
\begin{eqnarray}
  \frac{E_{{x}_1}}{k_{{x}_1}}\sigma_{R}^{x_1} & = & \int d\textbf{r}_{{x}_1}d\textbf{r}_{{x}_2} |\hat{S}_{b}(\textbf{r}_{{x}_1}, \textbf{r}_{{x}_2})|^{2}\label{psi4bapp-13}\\
  & & \qquad\times|\chi^{(+)}_{{x}_2}(\textbf{r}_{x_2})|^2
W(\textbf{r}_{{x}_1})|\chi^{(+)}_{{x}_1}(\textbf{r}_{x_1})|^2 , \nonumber
\end{eqnarray}
\begin{eqnarray}
  \frac{E_{{x}_2}}{k_{{x}_2}}\sigma_{R}^{x_2} & = & \int d\textbf{r}_{{x}_1}d\textbf{r}_{{x}_2} |\hat{S}_{b}(\textbf{r}_{{x}_1}, \textbf{r}_{{x}_2})|^{2}\label{psi4bapp2-13}\\
  & & \qquad\times |\chi^{(+)}_{{x}_1}(\textbf{r}_{x_1})|^2
W(\textbf{r}_{{x}_2})|\chi^{(+)}_{{x}_2}(\textbf{r}_{x_2})|^2 . \nonumber
\end{eqnarray}

The three-body reaction cross section $\sigma_{R}^{3B}$ corresponds to a new type of absorption where both fragments interacts with the target simultaneously ($x_1$ + $x_2$ + A).  To our knowledge the formulation for such a process has not yet been explored in the literature for the calculation of reaction cross sections. 

\section{Conclusions}
We summarized a few works aimed at extending the formalism developed by Hussein and McVoy \cite{HM1985} on Inclusive Non-Elastic Breakup (INEB) reactions. The main application of the formalism is to obtain neutron absorption cross sections by radioactive nuclei in inverse kinematics. First, we  have considered the case of a radioactive projectile incident on a two-cluster target, such as the deuteron. Applications have been done to extract neutron capture cross sections on neutron poisons based on the results of Ref. \cite{Bert19}. Further, an assessment of the statistical CDCC theory developed in Ref. \cite{BDH14} has lead to an additional method using random matrix theory. 

The case of INEB reactions two-projectile cluster (or fragment) reacting with a two-target cluster has also been discussed  \cite{Hus20}. A schematic formulation  was also introduced for the case of INEB reactions of three cluster projectiles with two cluster targets. All these ideas and theoretical methods have been elaborated in collaboration with M.S. Hussein along many years and decades. His enthusiasm and openness to solve problems in physics was a motivation for many of his collaborators, including ourselves.

\section*{Acknowledgements} 
BVC thanks P. Descouvemont for discussions and his hospitality at the Université Libre de Bruxelles where part of this work was developed.
CAB acknowledges support from U.S. DOE Grant No. DE-FG02-08ER41533as well as funding contributed through the LANL Collaborative Research Program by the Texas A\&M System National Laboratory Office and Los Alamos National Laboratory. BVC acknowledges support from grant 2017/05660-0 of the S\~ao Paulo Research Foundation (FAPESP), grant 306433/2017-6 of the CNPq and the INCT-FNA project 464898/2014-5.


\end{document}